# Hydrodynamics of Binary Coalescence.
# I. Polytropes with Stiff Equations of State.

Frederic A. Rasio[1]

Institute for Advanced Study, Olden Lane, Princeton, NJ 08540

and

Stuart L. Shapiro[2]

Center for Radiophysics and Space Research, Cornell University, Ithaca, NY 14853

## ABSTRACT

We have performed a series of three-dimensional hydrodynamic calculations of binary coalescence using the smoothed particle hydrodynamics (SPH) method. The initial conditions are exact polytropic equilibrium configurations on the verge of dynamical instability. We consider synchronized equilibria only, focusing on the effects of varying the compressibility of the fluid and the binary mass ratio. We concentrate here on stiff equations of state, with adiabatic exponents $\Gamma > 5/3$, and we assume that the polytropic constants ($K \equiv P/\rho^{\Gamma}$) are the same for both components. These conditions apply well to models of neutron star binaries. Accordingly, we discuss our results in the context of the LIGO project, and we calculate the emission of gravitational radiation in the quadrupole approximation. The fully nonlinear development of the instability is followed using SPH until a new equilibrium configuration is reached by the system. We find that the properties of this final configuration depend sensitively on both the compressibility and mass ratio. An *axisymmetric* merged configuration is always produced when the adiabatic exponent $\Gamma \lesssim 2.3$. As a consequence, the emission of gravitational radiation shuts off abruptly right after the onset of dynamical instability. In contrast, *triaxial* merged configurations are obtained when $\Gamma \gtrsim 2.3$, and the system continues to emit gravitational waves after the final coalescence. Systems with mass ratios $q \neq 1$ typically become dynamically unstable before the onset of mass transfer. Stable mass transfer from one neutron star to another in a close binary is therefore probably ruled out. For a mass ratio $q \lesssim 0.5$, however, dynamical mass transfer can temporarily retard the coalescence by causing a rapid reexpansion of the binary into a new, slightly eccentric but dynamically stable orbit. The maximum amplitude $h_{max}$ and peak luminosity $L_{max}$ of the gravitational waves emitted during the final coalescence are nearly independent of $\Gamma$, but depend sensitively on the mass ratio $q$. The approximate scalings we find are $h_{max} \propto q^2$ and $L_{max} \propto q^6$ for $q$ close to unity. These are much steeper dependences than would be expected for a system containing two point masses, where $h \propto q$ and $L \propto q^2(1 + q)$.

---

[1]Hubble Fellow.

[2]Departments of Astronomy and Physics, Cornell University.

*Subject headings:* hydrodynamics – instabilities – stars: neutron – stars: rotation – stars: binaries: close – radiation mechanisms: gravitational

## 1. INTRODUCTION

The coalescence and merging of two stars into a single object is the almost inevitable end-point of close binary evolution. Dissipation mechanisms such as the emission of gravitational radiation or fluid viscosity are always present and drive the orbital decay on a time scale that is almost always shorter than the internal evolution time scale of each star. It was only recently realized that the final stage of this orbital decay can often be hydrodynamic in nature, with the final merging of the two stars taking place on a time scale comparable to the rotation period (Rasio & Shapiro 1992, hereafter RS1; Lai, Rasio, & Shapiro 1993a,b, 1994a,b,c, hereafter LRS1–5 or collectively LRS). This is because *global hydrodynamic instabilities* can drive the binary system to rapid coalescence once the tidal interaction between the two components becomes sufficiently strong. These instabilities are caused by purely Newtonian tidal effects, which can lead to a steepening of the effective interaction potential in a close binary system.

Binary coalescence has been associated with a number of astrophysical phenomena of great current interest. Coalescing *white-dwarf* binaries are now generally thought to be the progenitors of type Ia supernovae (Iben & Tutukov 1984; Yungelson et al. 1994). They are also promising sources of low-frequency gravitational waves that should be easily detectable by future space-based interferometers (Evans, Iben, & Smarr 1987). In addition to producing supernovae, the coalescence of two white dwarfs may also lead in certain cases to the formation by gravitational collapse of an isolated millisecond pulsar (Chen & Leonard 1993) or the formation of blue subdwarf stars in globular clusters (Bailyn 1993). In the case of coalescing magnetized white dwarfs, a neutron star with extremely high magnetic field may form, and this scenario been proposed as a source of $\gamma$-ray bursts (Usov 1992). Contact *main-sequence-star* binaries are directly observed as Algol and W UMa systems (see Rucinski 1992 for a recent review). These systems are the likely progenitors of blue stragglers in globular clusters (Mateo et al. 1990).

White dwarfs and main-sequence stars typically have softer equations of state and less homogeneous density profiles than those of neutron stars. In addition, the mass-radius relation for main-sequence stars is fundamentally different from that of degenerate stars, leading to different binary properties when the mass ratio $q \neq 1$. Calculations focusing on white-dwarf and main-sequence-star models are therefore presented in a separate paper (Rasio & Shapiro 1994, hereafter Paper II). In this paper, we focus on systems containing degenerate stars with a stiff equation of state. Clearly, the most important astrophysical application is to the coalescence of neutron-star binaries, but our results can also be applied to hypothetical binary systems containing other degenerate objects with stiff equations of state, such as brown dwarfs, planets or planetesimals (cf. LRS4, §4).

Coalescing neutron-star binaries have long been recognized as important sources of gravitational radiation detectable by the new generation of laser interferometers such as LIGO (Clark 1979; Thorne 1987; Abramovici et al. 1992; Cutler et al. 1992). Statistical arguments based on the observed local population of binary pulsars with probable neutron star companions lead to an estimate of the rate of neutron star binary coalescence in the Universe of order $10^{-7} \, \mathrm{yr}^{-1} \, \mathrm{Mpc}^{-3}$ (Narayan, Piran & Shemi 1991; Phinney 1991). Finn & Chernoff (1993) estimate that an advanced LIGO detector could observe about 70 events per year. In addition to providing a major confirmation of Einstein's theory of general relativity, the detection of gravitational waves from coalescing binaries at cosmological distances could provide the first accurate measurement of the Universe's Hubble constant and mean density (Schutz 1986; Cutler et al. 1992; Chernoff & Finn 1993). Coalescing neutron stars are also at the basis of numerous models of $\gamma$-ray bursters (see Narayan, Paczyński, & Piran 1992, and references therein).

Recent calculations of the gravitational radiation waveforms from coalescing neutron star binaries have focused on the signal emitted during the last few thousand orbits, as the frequency sweeps upward from about 10 Hz to 1000 Hz. The waveforms in this regime can be calculated fairly accurately by performing high-order post-Newtonian expansions of the equations of motion for two *point masses* (Lincoln & Will 1990; Junker & Schäfer 1992; Kidder, Will, & Wiseman 1992). High accuracy is essential here because the observed signal will be matched against theoretical templates. Since the templates must cover $\gtrsim 10^3$ orbits, a fractional error as small as $10^{-3}$ can prevent detection. When, at the end of the inspiral, the binary separation becomes comparable to the stellar radii, hydrodynamic effects become important and the character of the waveforms will change. Special purpose narrow-band detectors that can sweep up frequency in real time will be used to try to catch the final few cycles of gravitational radiation (Meers 1988; Strain & Meers 1991). In this terminal phase of the coalescence, the waveforms should contain information not just about the effects of relativity, but also about the radius and internal structure of a neutron star. Since the masses and spins of the two stars, as well as the orbital parameters, can be determined very accurately from the lower-frequency inspiral waveform, the measurement of a single quantity such as the peak amplitude or frequency of the signal should suffice to determine the neutron star radius and place severe constraints on nuclear equations of state (Cutler et al. 1992). Such a measurement requires theoretical knowledge about all relevant hydrodynamic processes.

Two regimes can be distinguished in which different hydrodynamic processes take place. The first regime corresponds to the 10 or so orbits preceding the moment when the surfaces of the two stars first come into contact. In this regime, the two stars are still approaching each other in a quasi-static manner, but the tidal effects are very large. The second regime corresponds to the subsequent merging of the two stars into a single object. This involves very large departures from hydrostatic equilibrium, including mass shedding and shocks, and can be studied only by means of fully three-dimensional hydrodynamic computations. Such three-dimensional computations have been attempted only recently, first by Nakamura & Oohara (1991 and references therein; Oohara & Nakamura 1992) using traditional finite-difference techniques of numerical hydrodynamics, and later by Rasio & Shapiro (1992, hereafter RS1), Davies et al. (1993), and Zhuge, Centrella, & McMillan (1993) using the smoothed particle hydrodynamics (SPH) method.

In the first regime, the evolution of the system can still be described fairly accurately by a sequence of near-equilibrium fluid configurations. Such a description has been adopted in the recent work by LRS (see especially LRS2 and LRS3). Since neutron stars are not very compressible, the equilibrium configurations are not very centrally condensed and the usual Roche model for close binaries (e.g., Kopal 1959) does not apply. Instead, the hydrostatic equilibrium equation must be solved in three dimensions for the structure of the system. LRS use an approximate energy variational method to study *analytically* the hydrostatic equilibrium and stability properties of Newtonian binary systems obeying polytropic equations of state. In particular, they have constructed the compressible generalizations of all the classical solutions for binaries containing incompressible ellipsoidal components (Chandrasekhar 1969).

An important result found by LRS is that binary configurations containing a sufficiently incompressible fluid can become *dynamically unstable*. Close binaries containing neutron stars with stiff equations of state ($\Gamma \gtrsim 2$) should be particularly susceptible to these instabilities. As the dynamical stability limit is approached, the secular orbital decay driven by gravitational wave emission can be dramatically accelerated (LRS2, LRS3). The two stars then plunge rapidly toward each other, and merge together into a single object after just a few orbits.

This dynamical instability was indeed identified in RS1, where we calculated the evolution of equilibrium configurations containing two identical polytropes with $\Gamma = 2$. It was found that when the separation between the two stars becomes less than about three times the stellar radius, the orbit is unstable and the stars merge in just a few orbital periods. For larger binary separations, the system could be evolved dynamically for many orbital periods without showing any sign of orbital evolution. This paper is the first of a series where we examine in detail the effects of dynamical instabilities on a variety of close binary systems. We will concentrate on polytropes in synchronized binaries at first, varying the compressibility, masses and radii over wide ranges. More realistic stellar models for specific types of binary systems of particular interest will be considered later. For models of neutron star binaries, we will also consider nonsynchronized equilibrium configurations, which may be more likely in real systems.

In §2 we review our numerical method, and, in particular, we describe how to construct equilibrium configurations for close binaries using SPH. In §3 we present our results for polytropes with stiff equations of state, first for two identical stars (§3.1), and then for systems with $q \neq 1$ (§3.2). Further discussions of these results and speculations on future results are given in §4.

## 2. NUMERICAL METHOD

### 2.1. The Smoothed Particle Hydrodynamics Code

The smoothed particle hydrodynamics (SPH) method has been used for the calculations presented here. SPH is a Lagrangian method that was introduced specifically to deal with astrophysical problems involving self-gravitating fluids moving freely in three dimensions. The key idea is to calculate the pressure gradient forces by kernel estimation, directly from the particle positions, rather than by finite differencing on a grid, as is done in more traditional Lagrangian methods such as PIC. This idea was originally introduced by Lucy (1977) and Gingold & Monaghan (1977), who applied it to the calculation of dynamical fission instabilities in rapidly rotating stars. Since then, a wide variety of astrophysical fluid dynamics problems have been tackled using SPH (see Monaghan 1992 for a recent review). In the past few years, these have included galaxy formation (Katz 1992), star formation (Monaghan and Lattanzio 1991), solar system formation (Boss et al. 1992), supernova explosions (Herant & Benz 1992), tidal disruption of stars by massive black holes (Laguna et al. 1993), stellar collisions (Lai, Rasio, & Shapiro 1993c), as well as binary coalescence (Rasio & Shapiro 1992).

Our SPH code was developed originally by Rasio (1991) specifically for the study of hydrodynamic stellar interactions (Rasio & Shapiro 1991, 1992). The implementation of the SPH scheme is similar to that adopted by Hernquist & Katz (1989), but the gravitational field is calculated using a fast grid-based FFT solver. The neighbor searching in the code is performed using a new grid-based algorithm. Specifically, a multigrid hierarchical version of the linked-list method usually adopted in $P^3M$ particle codes (Hockney & Eastwood 1988) has been developed. The improved algorithm is extremely efficient, even for very nonuniform distributions of particles, provided that one is careful to fine-tune the ratio $L_g/\langle h_i \rangle_g$ of grid separation $L_g$ to the average SPH smoothing length $\langle h_i \rangle_g$ for that grid. Other details about the implementation, as well as a number of test-bed calculations using our SPH code for binary systems were presented in RS1. All calculations shown here were done using $N = 4 \times 10^4$ SPH particles, and each particle interacting with a nearly constant number of neigbors $N_N \approx 64$. The gravitational potential is calculated by FFT on a $256^3$ grid. The grid covers typically the entire system, but in cases where a small fraction of the mass is ejected to large distances we often prevent the grid boundary from expanding beyond a certain point, allowing up to 2% of the total mass to lie outside the grid. Forces on SPH particles outside the grid are then calculated by a multipole expansion.

We use a constant number density of SPH particles with varying particle masses to construct the initial conditions. This is in order to maintain good spatial resolution and mass resolution near the stellar surfaces, which is particularly important for problems involving tidal interactions and mass transfer in close binaries. We find that the SPH method is particularly well adapted to nearly incompressible fluids (high values of $\Gamma$), where the density drops sharply at the stellar surfaces. Such sharply defined surfaces would be very difficult to maintain accurately using a grid-based method of numerical hydrodynamics.

## 2.2. Conventions and Choice of Units

Throughout this paper, numerical results are given in units where $G = M = R = 1$, where $M$ and $R$ are the mass and radius of the *unperturbed* (spherical) primary (i.e., the more massive of the two stars). The units of time, velocity, and density are then

$$
\begin{align}
t_o &= 0.073 \, \text{ms} \times R_{10}^{3/2} M_{1.4}^{-1/2} \tag{1} \\
v_o &= 0.46 \, c \times R_{10}^{-1/2} M_{1.4}^{1/2} \tag{2} \\
\rho_o &= 2.8 \times 10^{15} \, \text{g cm}^{-3} \times R_{10}^{-3} M_{1.4}, \tag{3}
\end{align}
$$

where $M_{1.4}$ is the mass of the primary neutron star in units of $1.4 M_\odot$, $R_{10}$ is its radius in units of 10 km, and $c$ is the speed of light.

The mass of the secondary is $M' \leq M$. We denote the mass ratio by $q = M'/M \leq 1$. The equilibrium radius $R'$ of the secondary is calculated assuming *constant specific entropy throughout the system*, i.e., we use the same polytropic constant $K = P/\rho^\Gamma = K' = P'/\rho'^\Gamma$ for both components when constructing the initial condition. This assumption is appropriate for cold degenerate stars. It gives the mass-radius relation

$$
\frac{R}{R'} = \left( \frac{M}{M'} \right)^{(\Gamma-2)/(3\Gamma-4)}. \tag{4}
$$

Thus $R = $ constant for $\Gamma = 2$, while $R \propto M^{1/5}$ for $\Gamma = 3$.

## 2.3. Constructing Equilibrium Solutions

In addition to its normal use for dynamical calculations, our SPH code can also be used to construct self-gravitating, hydrostatic equilibrium configurations in three dimensions.

For *synchronized* binaries, this is done by adding a linear friction term $-\mathbf{v}/t_{relax}$ to the Euler equation of motion in the corotating frame. This forces the system to relax to a minimum-energy state. We use $t_{relax} = 1$ in our units, which makes the damping of unwanted oscillations nearly critical and optimizes the computation time to converge toward an equilibrium (RS1). Initially, two spherical polytropes are placed at a separation $r$. While the relaxation takes place, the particle entropies are maintained constant and their positions are continuously adjusted (by a simple uniform translation along the binary axis) so that the separation between the two centers of mass remains equal to its initial value $r$. Simultaneously, the angular velocity $\Omega$ defining the corotating frame is continuously updated so that the net centrifugal and gravitational accelerations of the two centers of mass cancel exactly [3].

The separation $r$ between the centers of mass can either be kept strictly constant or allowed to drift very slowly (on a time scale much longer than $t_{relax}$) so that an entire equilibrium *sequence* is constructed in a single integration. In practice we let $r(t) = r(0) - t/t_{scan}$, where $r(0)$ corresponds to well separated components and $t_{scan} = 100$ in our units.

Very accurate equilibrium solutions can be constructed using this relaxation technique, with the virial theorem satisfied to an accuracy of about one part in $10^3$ (Fig. 1) and excellent agreement found with the analytic solutions of LRS (Fig. 2). Other detailed comparisons with analytic results were also presented in RS1, LRS1 and LRS4.

Note that for *nonsynchronized* binaries, the relaxation technique described above to construct equilibrium configurations does not work since one cannot find a reference frame in which the fluid velocity is zero. Instead, one must determine *self-consistently* the velocity field as part of the equilibrium solution. This makes numerical calculations for nonsynchronized binaries considerably more complicated. For polytropes with sufficiently stiff equations of state, however, the velocity field in a nonsynchronized binary (or at least its divergence-free part) could be approximated by that of two Darwin-Riemann ellipsoids, as described in LRS3 and LRS4. In a future paper, we intend to apply the results of LRS to the construction of initial conditions for the coalescence of nonsynchronized binaries (see §4.4). Of particular interest for modeling neutron star binaries are the irrotational Darwin-Riemann ellipsoids, which apply when the viscous time scale is always longer than the orbital decay time scale (Kochanek 1992; LRS3).

## 3. RESULTS

### 3.1. Binaries Containing Two Identical Components

#### 3.1.1. Motivation for Using $\Gamma = 3$

In RS1 we presented our first calculation of binary coalescence for two identical polytropes with $\Gamma = 2$. We used as initial condition an equilibrium configuration just past the dynamical stability limit, which we identified at a binary separation $r/R \approx 3$ (corresponding to a slightly detached binary configuration).

---

[3] This is somewhat different from the procedure described in RS1, where we kept the total angular momentum constant during the relaxation while allowing $r$ to evolve. The disadvantage of that method was that, in addition to internal pulsations of period $\approx 1$, much longer-period epicyclic oscillations also had to damp before an equilibrium was reached, requiring considerably longer integrations to converge to an equilibrium. In addition, we find it more convenient to be able to fix the binary separation $r$ *a priori*.

We have now repeated this calculation with higher numerical accuracy, and using a slightly stiffer equation of state, corresponding to $\Gamma = 3$. The physical motivation for increasing $\Gamma$ was discussed in §6 of RS1: Although the ratio $T/|W|$ of kinetic to gravitational binding energy can reach values well in excess of the minimum necessary for sustaining a triaxial configuration ($(T/|W|)_{crit} = 0.14$), such triaxial configurations are violently unstable to mass shedding when the adiabatic index $\Gamma < 2.25$ (corresponding to a polytropic index $n > 0.8$; cf. James 1964). Indeed for $\Gamma = 2$ we found in RS1 that, after a phase of dynamical mass shedding, the system evolved to a new, *axisymmetric* equilibrium configuration with $T/|W| < 0.14$.

Since a stationary, axisymmetric configuration does not radiate gravitational waves, this dynamical evolution toward axisymmetry is characterized by an abrupt disappearance of the gravitational radiation signal at the end of the coalescence. In contrast, we pointed out in RS1 that the coalescence of two neutron stars with a slightly stiffer equation of state could result in the formation of a rotating, *triaxial* configuration (basically a compressible Jacobi ellipsoid; cf. LRS1). In this case the system would continue to emit gravitational radiation after the coalescence, although the amplitude of the waves would still drop abruptly to a level somewhat below the maximum reached just after the onset of dynamical instability. These lower-amplitude waves could be relatively easy to detect since they could persist for a large number of rotation periods. Ultimately, of course, gravitational radiation reaction forces will drive the system to a nonradiating configuration, either axisymmetric or still ellipsoidal but stationary (i.e., a compressible Dedekind ellipsoid; see Miller 1974, LRS1, and LRS5).

### 3.1.2.   The Dynamical Stability Limit

Our new study begins with a careful determination of the dynamical stability limit along the binary equilibrium sequence with $\Gamma = 3$. A sequence of equilibrium configurations with varying binary separation $r$ was constructed using the method described in §2.2. Several configurations near $r = 3$ were used as initial conditions for fully dynamical SPH calculations. On a time scale of just a few orbital periods, an unstable configuration evolves toward rapid merging of the two components, whereas stable configurations exhibit only small, constant-amplitude epicyclic oscillations about their circular orbit.

The results are illustrated in Figure 3, which shows the time evolution of the binary separation $r$ for three different dynamical calculations. Clearly, the system with initial separation $r = 2.95$ is unstable. After a brief excursion toward larger $r$ (basically half an epicycle), the system enters a dynamical coalescence, with $r$ decreasing monotonically. The system with $r = 3.1$ is clearly stable. There are only very small-amplitude ($\Delta r/r \lesssim 10^{-3}$), sinusoidal epicyclic oscillations. This merely reflects the impossibility of setting up a perfectly circular orbit numerically. The epicyclic frequency is very close to the orbital period ($P_{orb} = 24.2$), indicating that the orbit is very close to Keplerian. In contrast, the configuration with initial $r = 3.0$ is still on the stable side, but very close to the dynamical stability limit, and deviations from a Keplerian orbit are clearly apparent. The period of the epicyclic oscillations ($P_{osc} \approx 32$) is significantly longer than the orbital period ($P_{orb} = 22.5$). Indeed, if $r$ were precisely at the dynamical stability limit, the period of small epicyclic oscillations would be formally infinite. The oscillations also have larger amplitude ($\Delta r/r \approx 0.003$), and are markedly nonsinusoidal. This is to be expected as the stability limit is approached, since the amplification of any small departure from strict circular equilibrium in the initial data becomes stronger. Note also that the half epicycles at $r \gtrsim 3$ are shorter than those at $r \lesssim 3$ since the instantaneous restoring force for small radial perturbations decreases rapidly as $r$ approaches the stability limit from above. The most probable location of the dynamical stability limit, based on these numerical results, is between the two dashed lines in Figure 3. It corresponds to a binary separation $r \approx 2.97$, with a numerical error less than one percent. For comparison, the ellipsoidal, quasi-analytic calculations of LRS4 predict the onset of dynamical instability at $r \gtrsim 2.7$ for $\Gamma = 3$ ($n = 0.5$; cf. Table 13 of LRS1). The agreement within $\lesssim 10\%$ is quite satisfactory, considering the enormously greater computational cost of obtaining a more accurate estimate using the fully three-dimensional SPH calculations (several hours of supercomputer time for the SPH calculations versus a few seconds on a workstation using the methods of LRS).

### 3.1.3. Dynamical Evolution of an Unstable System

Quite apart from its better quantitative accuracy, the great advantage of using SPH for studying the equilibrium and stability properties of binary configurations is that the nonlinear evolution and final fate of unstable systems can then be readily determined. One simply continues the integration for an additional number of dynamical times until a new equilibrium or steady-state configuration is reached. The dynamical evolution of the system with initial binary separation $r = 2.95$, just past the dynamical stability limit (cf. Fig. 3), is illustrated in Figures 4–7. The complete calculation, performed with $N = 20000$ SPH particles per star, each particle interacting with $N_N \approx 64$ neighbors at all times, and a $256^3$ grid for the gravitational field, took about 60 CPU hours on an IBM ES-9000 supercomputer.

The radial infall velocity $v_r = \dot{r}$ (Fig. 4) remains quite small for about one orbital period, but then increases rapidly in magnitude to reach a maximum value $|v_r| \approx 0.08$ when $r \approx 2.5$. For comparison, consider the radial infall velocity induced by the gravitational radiation reaction on two identical neutron stars separated by the same distance and modeled as point masses. Using the quadrupole formula, and defining $\dot{r}_{GW} \equiv \dot{E}_{GW}/(dE/dr)$, we get $|\dot{r}_{GW}|/(GM/R)^{1/2} \approx 0.015 \times M_{1.4}^3 R_{10}^{-3}$, where $M_{1.4}$ is the mass of the neutron star in units of $1.4 M_\odot$ and $R_{10}$ is its radius in units of $10\,\mathrm{km}$. Clearly, the hydrodynamic instability dominates over the dissipative effects of the gravitational radiation for typical neutron stars. This is in agreement with the qualitative predictions obtained in LRS2 and LRS3.

The long-term evolution of the system is shown in Figure 5. The instability leads to the coalescence of the two stars on a time scale comparable to the orbital period. For neutron stars, the initial binary separation is $r \approx 28\,\mathrm{km}\,R_{10}$ and the initial orbital period is $P_{orb} \approx 1.5\,\mathrm{ms}\,R_{10}^{3/2} M_{1.4}^{-1/2}$.

The entire evolution shown in Figure 5 then takes about $7 \, \mathrm{ms} \, R_{10}^{3/2} M_{1.4}^{-1/2}$. Projections of all SPH particles into the orbital plane are shown at various times. The orbital rotation is counterclockwise. Initially, the evolution of the system is qualitatively similar to what we found in RS1. During the initial, linear stage of the instability ($t \approx 0\text{--}30$ in Fig. 1a–d), the two stars approach each other and come into contact after just about one orbital revolution. In the corotating frame of the binary, the relative velocity remains very subsonic (cf. Fig. 4), so that the evolution is adiabatic at this stage. This is in sharp contrast to the case of a head-on collision between two stars on a free-fall, radial orbit, where shocks are very important for the dynamics (RS1, §4). Here the stars are constantly being held back by a (slowly receding) centrifugal barrier, and the merging, although dynamical, is much more gentle. This can be seen clearly in Figure 6, where we show the time evolution of the various energies in the system. Notice that there are only small variations in time. The system remains close to virial equilibrium throughout its evolution. Contrast this to the case of a head-on collision, where there is a large transfer of kinetic to internal energy through recoil shocks, and practically no kinetic energy left at all in the final configuration (see Fig. 6 of RS1).

### 3.1.4. Mass Shedding and the Formation of the Outer Halo

After a little over two orbital revolutions, when even the innermost cores of the two stars have merged (Fig. 5f), mass shedding sets in rather abruptly. Material is ejected from the central object through the outer Lagrangian points of the effective potential and spirals out rapidly (Fig. 5g). In the final stage, the spiral arms widen and merge together in the central region (Fig. 5h–5j), but not outside (see below). The relative radial velocities of neighboring arms as they merge are supersonic, leading to shock-heating and dissipation. As a result, a hot, nearly axisymmetric rotating halo forms around the central dense core. No measurable amount of mass escapes from the system. The final, equilibrium structure of the halo is very similar to what we described in detail in RS1 for the $\Gamma = 2$ case. It contains about 20% of the total mass. The angular velocity decreases as a power-law $\Omega \propto r_{cyl}^{-\nu}$ where $r_{cyl} = (x^2 + y^2)^{1/2}$ and $\nu \lesssim 2$ (cf. Fig. 15 of RS1).

The outer spiral arms, although transient in nature, form a very extended coherent structure that can subsist for a large number of central dynamical times (Fig. 7). The possibility that the spiral arms formed during the coalescence of two neutron stars could fragment through a sausage instability (Chandrasekhar 1961, Chap. 12), with the fragments later exploding on their $\beta$-decay time scale, was discussed recently by Colpi & Rasio (1994) in the context of $\gamma$-ray bursts. Such fragmentation may indeed be visible in the outer parts of the spiral arms at late times (see Fig. 7b), although the size of the fragments is at the limit of our spatial resolution. Comparing the large-scale structure of the halo at similar times in the calculations with $\Gamma = 2$ and $\Gamma = 3$ (e.g., Fig. 7b here with Fig. 13 in RS1), we see that the outer spiral arms remain narrow and well-defined for a much longer time when the fluid is less compressible. This is a simple consequence of the smaller specific enthalpy of the less compressible fluid. For a strictly incompressible fluid, the spiral arms would never merge, but would instead become thinner and thinner as they wind up.

### *3.1.5. Properties of the Central Merged Configuration*

Our most significant new result concerns the structure of the central core. As we predicted in RS1, it has a *triaxial* configuration. Since the rotation is nearly uniform (see below), its structure is basically that of a *compressible Jacobi ellipsoid* (cf. LRS1).

This can be seen directly by close inspection of Figure 5 (especially Fig. 5k), although the deviation from axisymmetry in the spatial distribution of SPH particles is quite small. A blow-up of the central region at the end of our calculation is shown in Figure 8b (bottom frame). Here contours of equal density are shown in the equatorial plane. The axis ratio $a_2/a_1$ varies between about 0.83 near the center and 0.77 for the outermost contour shown. For comparison, the top frame shows the initial (marginally unstable) binary configuration. Similar results for $\Gamma = 5/3$ and $\Gamma = 10$ are also shown in Figure 8 (see below). Note that the ellipsoidal core has a well-defined, sharp surface where the density decreases abruptly (Fig. 9). In our units the central density $\rho_c \approx 0.4$, giving $\rho_c \approx 1.0 \times 10^{15} \, \text{g cm}^{-3} \, M_{1.4} \, R_{10}^{-3}$. This is nearly identical to the central density of the initial stellar model. The angular velocity of the fluid inside the core is very nearly constant (Fig. 9), with an average value of $\Omega_c = 0.48$ in our units. This corresponds to a rotation period $(2\pi/\Omega)_c \approx 1 \, \text{ms} \, M_{1.4}^{-1/2} \, R_{10}^{3/2}$ for the central ellipsoid.

### *3.1.6. Gravitational Wave Emission*

We have calculated the emission of gravitational radiation during the dynamical coalescence using the quadrupole approximation (see RS1, §2.1). Figure 10 shows the amplitudes of the two polarizations $h_+$ and $h_\times$ for an observer situated at a distance $r_O$ along the rotation axis,

$$r_O h_+(0,0) = \frac{G}{c^4}(\ddot{\text{I}}_{xx} - \ddot{\text{I}}_{yy}), \qquad r_O h_\times(0,0) = 2\frac{G}{c^4}\ddot{\text{I}}_{xy}, \tag{5}$$

where $\text{I}_{ij}$ is the reduced quadrupole moment. Clearly, these quantities are a direct measure of the deviation from axisymmetry. The amplitude drops abruptly after the onset of mass shedding, but then stabilizes to a nonzero value. The maximum amplitude (around $t = 35$) corresponds to $r_O h_{max} \approx 2.2/c^4$ in our units, or $h_{max} \approx 3 \times 10^{-21} \, M_{1.4}^2 \, R_{10}^{-1} \times (r_O/10 \, \text{Mpc})^{-1}$. The final amplitude $h_{final} \approx 0.1 \, h_{max}$, in contrast to $h_{final} = 0$ to high precision for $\Gamma = 2$ (RS1). Thus a simple determination of the absence or presence of persisting gravitational radiation after the coalescence (i.e., after the sudden drop in amplitude) could place an interesting constraint on the stiffness of the equation of state.

Waveforms emitted in a direction away from the rotation axis have smaller amplitudes. Because the orbital ($z = 0$) plane is a symmetry plane for the system, we have $\text{I}_{xz} = \text{I}_{yz} = 0$ and the polarization $h_\times$ of the radiation in any direction $(\theta, \phi)$ can be written in terms of the two polarizations emitted along the axis,

$$h_\times(\theta, \phi) = -h_+(0,0) \cos\theta \sin 2\phi + h_\times(0,0) \cos\theta \cos 2\phi \tag{6}$$

(cf. RS1, eqns. [24], [25], and [47], with $\ddot{\mathsf{i}}_{kz} = 0$). In the equatorial plane, the emission is minimum, with $h_\times(\pi/2, \phi) = 0$ by symmetry. The amplitude $h_+(\pi/2, \phi) = \ddot{\mathsf{i}}_{zz}/r_O \sim 10^{-3} h_{max}$ in our calculations. It is determined primarily by small-amplitude vertical pulsations of the fluid.

Note that the numerical value of $r_O h_{max}$ is nearly identical to what we found in RS1 for $\Gamma = 2$. Indeed, the tendency for $r_O h_{max}$ (when expressed in our Newtonian units) to remain constant independent of $\Gamma$ is confirmed by calculations we did for the two more extreme cases $\Gamma = 5/3$ and $\Gamma = 10$ (see §3.1.7 below and Fig. 10). When expressed in geometrized units, where $G = c = 1$, this important result can be written

$$\left(\frac{r_O}{M}\right) h_{max} \approx \text{constant} \times \left(\frac{M}{R}\right), \tag{7}$$

where the constant $\approx 2.2$, *independent of the details of the equation of state* (at least in the range $5/3 \lesssim \Gamma \lesssim 10$). Thus a measurement of $h_{max}$ for two identical neutron stars by LIGO would give the radius $R$ of a neutron star since the mass $M$ can be determined very accurately from the low-frequency inspiral waveform observed over the last few minutes prior to the final coalescence (Cutler et al. 1992).

The total luminosity of gravitational radiation is shown in Figure 11. There is a single broad peak, whose width is about equal to one orbital period. The total energy radiated in gravitational waves at the end of the calculation is $\Delta \mathcal{E}_{GW} \approx 8/c^5$ in our units, or $\Delta \mathcal{E}_{GW} \approx 10^{52}$ erg $M_{1.4}^{9/2} R_{10}^{-7/2}$, corresponding to an efficiency $\Delta \mathcal{E}_{GW}/Mc^2 \approx 3 \times 10^{-2} M_{1.4}^{7/2} R_{10}^{-7/2}$. This is not significantly different from what we found in RS1 for $\Gamma = 2$.

Comparing total integrated luminosities between different calculations may not be so useful since the origin of time is not precisely defined. However, the *peak luminosity* $L_{max}$ is well-defined and its value in our units appears to increase only slightly with increasing compressibility. Expressing again the results in geometrized units, we find

$$\left(\frac{L_{max}}{L_o}\right) \approx \text{constant} \times \left(\frac{M}{R}\right)^5, \tag{8}$$

where $L_o \equiv c^5/G = 3.6 \times 10^{59} \, \text{erg s}^{-1}$ and the constant $\approx 0.5$ independent of the equation of state. Given the very steep dependence on $M/R$, it is clear that $L_{max}$ should provide an even better measure of $R$ than $h_{max}$.

The principal differences found between the results of RS1 for $\Gamma = 2$ and our new results for $\Gamma = 3$ are summarized in Table 1.

### 3.1.7. More Extreme Cases

To better understand the effects of compressibility, and to confirm the conclusions reached by comparing our results for $\Gamma = 2$ and $\Gamma = 3$, we have performed similar dynamical coales-cence calculations for two more extreme cases, $\Gamma = 5/3$ (very compressible), and $\Gamma = 10$ (nearly incompressible). The results are summarized in Figures 8, 10, and 11.

Even for an adiabatic exponent as small as $\Gamma = 5/3$, equilibrium configurations for close binaries can become dynamically unstable. For $\Gamma \lesssim 2$, however, this occurs when the two components are already coalescing, i.e., the last stable configuration along an equilibrium sequence with decreasing angular momentum is a *contact configuration* (see Fig. 8a). It is only for an even more compressible fluid with $\Gamma \lesssim 1.5$ (polytropic index $n \gtrsim 2$) that the mass shedding limit is encountered first along a sequence of deeper and deeper contact equilibria. In such a case the ejection of mass through the outer Lagrangian points would be driven by dissipation, and mass loss would proceed on the secular orbital decay time scale (rather than dynamically).

The final merged configurations for $\Gamma < 2.25$ are always axisymmetric (cf. bottom of Fig. 7a and RS1, Fig. 16). The transition between core and halo is also much smoother in the more compressible cases, and the halos are more massive. For $\Gamma = 5/3$, about 30% of the total mass is ejected from the central region during the coalescence. This tendency for more compressible fluids to form more massive halos is easy to understand since they are more susceptible to mass shedding instabilities. Our numerical results for $\Gamma = 5/3$ have important implications for other astrophysical systems such as contact main-sequence-star (W UMa) binaries and double white-dwarf systems, and will be presented in detail elsewhere (Paper II; Rasio 1994).

For a nearly incompressible fluid [4] (Fig. 8c), the last stable configuration has a larger binary separation and is clearly detached. The final merged configuration is markedly triaxial, resembling a classical (incompressible) Jacobi ellipsoid. The axis ratio $a_2/a_1 \approx 0.6$ for the ellipsoid with $\Gamma = 10$ in Figure 8c.

If we calculate the values of the ratio $T/|W|$ for the central cores of the merged configurations we find $T/|W| \approx 0.14$ for $\Gamma = 3$ and $T/|W| \approx 0.16$ for $\Gamma = 10$. These values are very close to those calculated by Hachisu & Eriguchi (1982) for compressible Jacobi ellipsoids near the mass-shedding limit (see also LRS1, §3.3). This is indeed what we expect: During the initial phase of the coalescence, $T/|W|$ increases rapidly to a value well above the mass shedding limit $((T/|W|)_{max} \approx 0.20$, slightly less for smaller $\Gamma$). As much mass and angular momentum are removed from the central region as is needed to leave it with just the amount of rotation that it can sustain in a stable equilibrium state.

Figures 10 and 11 compare the gravitational radiation waveforms and luminosities emitted during the dynamical coalescence of three systems with widely different values of $\Gamma$. Decreasing $\Gamma$ clearly leads to a more abrupt extinction of the signal, with the amplitude decreasing all the way

---

[4]Objects as incompressible as $\Gamma = 10$ may in fact exist in Nature. In particular, *strange stars* (Witten 1984), made of bulk quark matter consisting of roughly equal numbers of up, down, and strange quarks, may have $\Gamma \gg 3$ (up to $\Gamma \approx 200$!) when the density is close to the nucleon-quark transition density (Haensel et al. 1986). Most properties of strange stars (in particular, their radii), are otherwise very similar to those of ordinary neutron stars. In fact, it is possible that all objects thought to be ordinary neutron stars are in fact strange stars (Alcock et al. 1986).

to zero for $\Gamma \lesssim 2$ (see also Fig. 17 of RS1 for $\Gamma = 2$).

## 3.2. Binaries with Nonidentical Components

### 3.2.1. The Problem of Mass Transfer

One can use exactly the same approach described in §3.1 to study the final coalescence of binaries with mass ratio $q \neq 1$. One essential difference, however, is that the equilibrium sequences terminate at a *Roche limit* which is always encountered before the surfaces of the two stars come into contact. Thus one has to worry about the possibility of either dynamical or secular mass transfer from one component to the other.

Clark & Eardley (1977) have suggested that secular, *stable* mass transfer from one neutron star to another could last for hundreds of orbital revolutions before the lighter star is tidally disrupted. Such an episode of stable mass transfer would be accompanied by a secular *increase* of the orbital separation. Thus if stable mass transfer could indeed occur, a characteristic "reversed chirp" would be observed in the gravitational wave signal at the end of the inspiral phase (Jaranowski & Krolak 1992). The problem has been reexamined more recently by Kochanek (1992) and Bildsten & Cutler (1992), who both argue against stable mass transfer on the basis that very large mass transfer rates and extreme mass ratios would be required.

The situation was clarified in LRS3, where it was pointed out that the existence of a Roche limit has in fact little importance for most neutron star binaries (except perhaps those containing a very low-mass neutron star). This is because for $\Gamma \gtrsim 2$, *dynamical instability always arises before the Roche limit* along an equilibrium sequence with decreasing $r$. Therefore, by the time mass transfer is triggered, the system is already in a state of dynamical coalescence. Nevertheless, as we will see in §3.2.3, even a very brief episode of dynamical mass transfer can have a profound influence on the qualitative evolution of the system.

### 3.2.2. Dynamical Coalescence of a Slightly Asymmetric System

We have performed complete dynamical calculations for two cases, one with mass ratio $q = 0.85$ and the other with $q = 0.5$. We used $\Gamma = 3$ in both cases. Results concerning the equilibrium properties of the configurations with $q = 0.85$ were presented in LRS4 (§5). The dynamical stability limit is at $r \approx 2.95$, whereas the Roche limit is at $r \approx 2.85$ (Recall that our units are based on the mass and radius of the *more massive* star). The value $q = 0.85$ is, given the presently available data, the most probable value of the mass ratio in the binary pulsar PSR 2303+46 (Thorsett et al. 1993) and represents the largest observed departure from $q = 1$ in any observed binary pulsar with likely neutron star companion. For comparison, $q = 1.386/1.442 = 0.96$ in PSR 1913+16 (Taylor & Weisberg 1989) and $q = 1.32/1.36 = 0.97$ in PSR 1534+12 (Wolszczan 1991).

The dynamical evolution of the system with $q = 0.85$ is shown in Figure 12. The initial configuration has $r = 2.9$. At $t = 0$, the star on the left has $M = R = 1$ and the star on the right has $M' = 0.85$ and $R' = (M'/M)^{1/5}R = 0.97$ (cf. §2.2). As before the orbital rotation is counterclockwise. The Roche limit is quickly reached while the system is still in the linear stage of growth of the instability. Dynamical mass transfer begins within the first orbital revolution. A relatively narrow accretion stream forms first (Fig. 12b), but it widens rapidly (Fig. 12c) as the orbital separation continues to decrease. Because of the proximity of the two components, the fluid acquires very little velocity as it slides down from the inner Lagrangian point to the surface of the other star. As a result, relative velocities of fluid particles remain largely subsonic and the coalescence proceeds quasi-adiabatically, just as in the $q = 1$ case. In fact, the mass transfer appears to have essentially no effect on the dynamical evolution in this case.

After about two orbital revolutions the smaller-mass star is tidally disrupted (Fig. 12d). Some material is ejected from around the outermost Lagrangian point and forms a single-arm spiral outflow (Fig. 12b). Because of this asymmetric ejection, the central core receives a small momentum kick. Its velocity with respect to the system center of mass at the end of our calculation is $v_c \approx 0.02$ in our units, or $v_c \approx 2.7 \times 10^3 \, \mathrm{km \, s^{-1}} \, M_{1.4}^{1/2} \, R_{10}^{-1/2}$. The fraction of the total mass ejected from the core is $\approx 0.13$, appreciably smaller than what we found in the $q = 1$ case.

The internal density and velocity profiles of the merged cores are practically indistinguishable in the $q = 1$ and $q = 0.85$ cases (see Fig. 9). However, the rearrangement of fluid elements from the original binary components is quite different, as illustrated in Figure 13. In the asymmetric case, a large central region of the more massive star remains little affected by the hydrodynamic interaction taking place outside. The less massive star has been tidally disrupted and its material spread over the surface of the more massive component. Note that, since the specific entropy is uniform throughout the entire system in our initial model, and since there is little dissipation during the dynamical evolution, the material coming from the less massive star is neither buoyant nor Rayleigh-Taylor unstable. In contrast, in a symmetric system, the two stars fill roughly two hemispheres separated by a meridional section in the final merged configuration.

### 3.2.3. More Extreme Mass Ratios

To our surprise, the dynamical evolution of a system with $q = 0.5$ turns out to be quite different. The results are illustrated in Figures 14 and 15. The initial binary separation is $r = 2.9$, so that the system is slightly past the dynamical stability limit but not yet at the Roche limit, exactly as in the $q = 0.85$ calculation[5].

---

[5]Note that the ratio $r_{dyn}/R$, where $r_{dyn}$ is the binary separation corresponding to the dynamical stability limit, is nearly independent of $q$ over the interval $0.5 \leq q \leq 1$. This is in agreement with the results of LRS4, who find $r_{dyn}/R = 2.68$ for $q = 1$ and $r_{dyn}/R = 2.71$ for $q = 0.5$ (see their Table 2). That both numbers are about 10% too small comes from the truncation of the tidal potential to quadrupole order in the analytic treatment of LRS.

At $t = 0$ (Fig. 14a), the star on the left has $M = R = 1$ and the star on the right has $M' = 0.5$ and $R' = 0.87$. After a brief episode of dynamical mass transfer (Fig. 14b), the orbital separation *increases* back to a value ($r_{final} \approx 3.1$) significantly larger than at $t = 0$, and a new *stable* binary configuration is reached. Although the new orbit is slightly eccentric ($e \approx 0.03$), there is no longer any mass transfer, even near pericenter passages (Fig. 14d).

Figure 15 shows the time evolution of the separation $r$, as well as the masses $M$ and $M'$ of the two components and the total entropy $S$ in the system. Component masses are calculated as described in Rasio & Shapiro (1991) and the total entropy is defined by equation (13) of RS1 with the arbitrary additive constant chosen so that $S = 0$ at $t = 0$. Here the (very brief) dynamical mass transfer is quite dissipative compared to the $q = 0.85$ case, and a significant increase in entropy is produced. The total mass transferred from $M'$ to $M$ is $\Delta M/M \approx 0.07$, nearly equal to the fractional increase in mean separation $\Delta r/r \approx (3.1 - 2.9)/2.9 \approx 0.07$ between the initial and final orbits. Note that this is *not* the simple result predicted for the Keplerian orbit of two point masses, where we would have $\Delta r/r \approx 2[(1 - q)/q]\Delta M/M \approx 2\Delta M/M$. Large deviations from Keplerian behavior are indeed expected here, since the spins and tidal distortions give large contributions to the total energy and angular momentum in the system (cf. LRS2, LRS3).

### 3.2.4. Gravitational Radiation Dependence on the Mass Ratio

Gravitational radiation waveforms for the two systems with $q < 1$ are shown in Figure 16. The waveform for $q = 0.85$ is qualitatively similar to the one shown in Figure 10 for $q = 1$, but the maximum amplitude is $r_O h_{max} \approx 1.6/c^4$ in our units, appreciably smaller.

The dependence of $h_{max}$ on the mass ratio $q$ is quite strong, and nontrivial, since it depends on the details of the hydrodynamics. If we simply assume that $h_{max} \propto q^\alpha$ and solve for $\alpha$ using the results for $q = 1$ and $q = 0.85$, we find $\alpha \approx 2$. Note that this is quite different from the scaling obtained for a detached binary system with a given binary separation. In particular, if we consider two point masses in a circular orbit with separation $r$ we have $h \propto \Omega^2 \mu r^2$, where $\Omega^2 = G(M + M')/r^3$ and $\mu = MM'/(M + M')$. At constant $r$, this gives $h \propto q$. This linear scaling is obeyed (only approximately, because of finite-size effects) by the wave amplitudes of the various systems at the *onset* of the dynamical instability. For determining the *maximum* amplitude, however, hydrodynamics plays an essential role. In a system with $q \neq 1$, the more massive star tends to play a far less active role in the hydrodynamics (in particular, it does not shed mass; compare, e.g., Fig. 5h and Fig.12d), and, as a result, there is a rapid suppression of the radiation efficiency as $q$ departs even slightly from unity.

The peak luminosity of gravitational radiation $L_{max} \approx 0.14/c^5$ for the $q = 0.85$ calculation. Setting $L_{max} \propto q^\beta$ we get $\beta \approx 6$. Again, this is a much steeper dependence than one would expect based on a simple point-mass estimate, which gives $L \propto q^2(1 + q)$ at constant $r$.

The approximate scalings $h_{max} \propto q^2$ and $L_{max} \propto q^6$ can apply only to systems with $q_{crit} \leq$

$q \leq 1$, where coalescence does occur immediately following the onset of dynamical instability and the amplitude of gravitational radiation drops abruptly after the maximum is reached. Based on our results to date we can only say that $0.5 < q_{crit} < 0.85$ for $\Gamma = 3$.

For $q = 0.5$, the radiation remains very close to that of a detached binary system throughout the dynamical evolution. Both the amplitude and the frequency of the waves increase only slightly after the onset of dynamical instability (Fig. 16). On a *secular* orbital decay time scale, determined by the radiation reaction, the system will be brought back to its dynamical stability limit and the type of evolution shown in Fig. 14 will repeat. This time scale can vary from about one to many orbital periods depending on the value of the relativistic parameter $R/M$ for the stars involved (see LRS4). To determine the long-term evolution of the system through many of these cycles and the corresponding waveforms would require the inclusion of the radiation reaction potential in the hydrodynamic calculations, which is beyond the scope of this paper.

## 4. DISCUSSION

### 4.1. Gravitational Wave Emission from Rotating Ellipsoids

Our results suggest that the coalescence of neutron stars with sufficiently stiff equations of state could very well lead to the formation of a more massive, rapidly rotating neutron star with a *triaxial* equilibrium configuration. Within our simple polytropic model, this would require $\Gamma \gtrsim 2.3$, which is satisfied by most realistic models for sufficiently massive neutron stars. For example, Table 3 of LRS3, based on the AV14+UVII nuclear equation of state of Wiringa, Fiks, & Fabronici (1988), predicts an effective adiabatic exponent $\Gamma > 2.3$ for all neutron star masses $M \gtrsim 1.0 M_\odot$.

The triaxial merged configuration can be modeled approximately as compressible Jacobi ellipsoids (LRS1, §4). The emission of gravitational waves by such rotating ellipsoids has peculiar characteristics. While the amplitude of the waves and total luminosity *decrease* as the ellipsoid radiates away angular momentum and evolves toward axisymmetry, the frequency of the waves actually *increases* (the ellipsoid spins *up*) during this process. This is because the angular velocity of rotation of both classical and compressible Jacobi ellipsoids is a decreasing function of angular momentum (see LRS1, Table 4 and Fig. 7).

The rotating ellipsoid is expected to be formed close to the mass shedding limit for such configurations, as we have seen in §3.1.6. Because the mass shedding limit along a compressible Jacobi sequence with $\Gamma < 10$ is so close to the bifurcation point where it branches off the Maclaurin sequence of axisymmetric equilibria (cf. Hachisu & Eriguchi 1982; LRS1), compressible Jacobi ellipsoids cannot be very elongated. For $\Gamma = 3$, we must have $a_2/a_1 \gtrsim 0.7$ (see LRS1, Tables 3 and 4), while our calculation for a binary with $q = 1$ indicates $a_2/a_1 \approx 0.8$. This limits the amplitude and luminosity of the gravitational radiation, since $h \propto [1 - (a_2/a_1)^2]$ and $L \propto [1 - (a_2/a_1)^2]^2$. At the same time, however, the reduced luminosity can make the emission last longer, making the object easier to detect by template matching techniques (see Thorne 1987). In contrast, the short ($\sim 5\,\mathrm{ms}$) burst of gravitational waves emitted during the final coalescence, although of much higher luminosity, may be quite difficult to detect (Cutler et al. 1992). Since the exact location of the mass shedding limit along a compressible Jacobi sequence depends so sensitively on $\Gamma$ (Hachisu & Eriguchi 1982), the total *duration* of the signal emitted by the rotating ellipsoid could provide an independent constraint on the neutron star equation of state.

The uniform rotation of the final ellipsoidal configuration is clearly a direct consequence of synchronization in the original binary system. However, even if the binary is nonsynchronized (see §4.4 below), triaxial merged configurations are still possible. These would be the compressible analogs of Riemann ellipsoids (LRS1, §5), in which internal fluid motions with uniform vorticity are present in the corotating frame of the ellipsoidal figure. In particular, a binary system containing two initially nonspinning stars could evolve into an *irrotational* Riemann ellipsoid. In fact, triaxial merged configurations may be *more likely* when the original binary is nonsynchronized. This is because no minimum value of $\Gamma$ is required for the existence of a compressible irrotational Riemann ellipsoid. In addition, mass shedding is expected to play a less important role. Indeed, it is generally found that relaxing the assumption of uniform rotation allows compressible equilibria with considerably higher values of $T/|W|$ to exist (see Bodenheimer & Ostriker 1973).

## 4.2. Measuring the Radius of a Neutron Star with LIGO

The most important parameter that enters into quantitative estimates of the gravitational wave emission during the final coalescence is the relativistic parameter $M/R$ for a neutron star. In particular, for two identical point masses we know that the wave amplitude obeys $(r_O/M)h \propto (M/R)$ and the total luminosity $L \propto (M/R)^5$. Thus one expects that any quantitative measurement of the emission near maximum should lead to a direct determination of the radius $R$, assuming that the mass $M$ has already been determined from the low-frequency inspiral waveform (Cutler et al. 1992). Most current nuclear equations of state for neutron stars give $R/M \approx 5\text{–}8$, with $R \approx 10\,\mathrm{km}$ nearly independent of the mass in the range $0.8 M_\odot \lesssim M \lesssim 1.5 M_\odot$ (see, e.g., Baym 1991 for a recent review).

However, as we have seen, the details of the hydrodynamics also enter into this determination. They introduce an explicit dependence of all wave properties on the internal structure of the stars (which we represent here by a single dimensionless parameter $\Gamma$), and on the mass ratio $q$. If relativistic effects were taken into account for the hydrodynamics itself, an additional, nontrivial dependence on $M/R$ would also be present. This can be written conceptually as

$$\left(\frac{r_O}{M}\right) h_{max} \equiv \mathcal{H}(q, \Gamma, M/R) \times \left(\frac{M}{R}\right) \tag{9}$$

$$\frac{L_{max}}{L_o} \equiv \mathcal{L}(q, \Gamma, M/R) \times \left(\frac{M}{R}\right)^5 \tag{10}$$

Combining the results of §3.1 and §3.2, we can write, in the limit where $M/R \rightarrow 0$ and for $q$ sufficiently close to unity,

$$\mathcal{H}(q, \Gamma, M/R) \approx 2.2q^2 \qquad \mathcal{L}(q, \Gamma, M/R) \approx 0.5q^6, \tag{11}$$

independent of $\Gamma$.

### 4.3. Dynamical Mass Transfer

It is clear from the results of §3.2.3 that mass transfer can have a retarding effect on the coalescence, even in the presence of a dynamical instability. This stabilizing tendency of mass transfer has two origins. First, for polytropes with $\Gamma > 2$, we have $(dR/dM)_K > 0$, i.e., the mass-losing star responds adiabatically by shrinking in size (cf. eq. [4]). Real neutron stars may have $(dR/dM)_S > 0$ or $(dR/dM)_S < 0$ depending on their mass, but for $0.3 \lesssim M/M_\odot \lesssim 2$, $R \approx 10$ km is nearly independent of mass (cf. LRS3, §4.1 and references therein) and the change of size may be negligible in determining the response of the system at the onset of mass transfer. The second and dominant stabilizing effect comes from the response of the orbital motion. Because of the close proximity between the stellar surfaces at the onset of mass transfer, the transferred material carries negligible specific angular momentum (cf. Fig. 14b). In that case, simple considerations based on the conservation of total angular momentum (see, e.g., Hut & Paczyński 1984; Benz et al. 1990) lead to the well-known result that mass transfer from the less massive to the more massive component leads to orbital expansion. Although this prediction is qualitatively in agreement with our numerical results, the simple arguments cannot be applied rigorously to close neutron star binaries since they are based on the Keplerian response of two point masses (cf. §3.2.3). Indeed, at the onset of mass transfer in a close neutron star binary, the spin angular momentum of each star can be comparable to the orbital angular momentum, and large deviations from Keplerian behavior are expected (see LRS3). The problem is further complicated by the loss of synchronization during dynamical mass transfer (since viscosity cannot act fast enough to maintain synchronization as the orbit readjusts on a dynamical time scale to the exchange of mass).

The situation can be completely opposite for systems containing degenerate stars with $\Gamma < 2$, such as white dwarfs. In those systems, the equilibrium configurations remain in general dynamically stable all the way down to the Roche limit, but the mass transfer itself can be dynamically unstable and drive the system to rapid coalescence. SPH calculations we did starting from the *dynamically stable* Roche limit configuration for two $n = 1.5$ ($\Gamma = 5/3$) polytropes with $q = 0.5$ and $K = K'$ demonstrate this type of evolution. The details will be presented in Paper II. A similar calculation for two white dwarfs using a somewhat more realistic degenerate equation of state was performed by Benz et al. (1990). Although their results are in qualitative agreement with ours, one may question their use of a nonequilibrium initial configuration.

### 4.4.  Directions for Future Work

Recent theoretical work suggests that the synchronization time in close neutron star binaries may remain always longer than the orbital decay time due to gravitational radiation (Kochanek 1992; Bildsten & Cutler 1992). In particular, Bildsten & Cutler (1992) show with simple dimensional arguments based on the standard weak-friction model of tidal interactions (Zahn 1977) that one would need an implausibly small value of the effective viscous time, $t_{visc} \sim R/c$, in order to obtain synchronization. In the opposite limiting regime where viscosity is completely negligible, the fluid circulation in the binary system is conserved during the orbital decay and the stars can be modeled approximately as Darwin-Riemann ellipsoids (Kochanek 1992; LRS3). Of particular importance are the irrotational Darwin-Riemann configurations, obtained when two initially non-spinning (or, in practice, slowly spinning) neutron stars evolve in the absence of significant viscosity. These configurations exhibit smaller deviations from point-mass Keplerian behavior at small $r$. An important result obtained in LRS3 is that even irrotational configurations for binary neutron stars with $\Gamma \gtrsim 2$ become dynamically unstable before the two stars come into contact. Thus the final coalescence of two neutron stars in a nonsynchronized binary system will also be driven by the hydrodynamic instability studied here for synchronized configurations.

The details of the hydrodynamics could be quite different, however. Because the two stars appear to be counter-spinning in the corotating frame of the binary, a vortex sheet with $\Delta v = |v_+ - v_-| \approx \Omega r$ will appear when the surfaces come into contact. Such a vortex sheet is Kelvin-Helmholtz unstable on all wavelengths and the hydrodynamics may therefore be quite difficult to model accurately given the limited spatial resolution of typical three-dimensional calculations. The breaking of the vortex sheet could generate a large turbulent viscosity so that the final configuration would no longer be irrotational. In numerical simulations, however, vorticity is likely to be generated mostly through spurious shear viscosity introduced by the spatial discretization. An additional difficulty of great concern is that nonsynchronized configurations evolving rapidly by gravitational radiation emission will tend to develop significant tidal lags, with the long axes of the two components becoming misaligned (see LRS5). This is a purely dynamical effect, present even if the viscosity is zero, but depending on the previous orbital evolution of the system. Thus the construction of initial conditions for hydrodynamic calculations of the coalescence of nonsynchronized neutron star binaries must incorporate the gravitational radiation reaction *self-consistently*.

Recent attempts at modeling the hydrodynamics of nonsynchronized, equal-mass binary co-alescence have been made by Shibata, Nakamura, & Oohara (1992a,b) and Davies et al. (1993). In both studies, the initial conditions consisted of two identical spinning *spheres* (polytropes with $\Gamma \approx 2$) placed on an inspiral trajectory calculated for two point masses. These initial conditions are neither in equilibrium, nor self-consistent, and the results may therefore not be physical. Somewhat more troubling is that, although they started from essentially identical initial conditions, the two groups find subsequent dynamical evolutions that are not even in qualitative agreement. Whereas Davies et al. (1993) obtain a final merged configuration that is axisymmetric (and with a large, uniformly rotating central core — a clear indication that vorticity was not conserved in the calculation), Shibata et al. (1992a,b) obtain a nonaxisymmetric final configuration with a double-core structure. Mass shedding through spiral arms is seen in the Davies et al. results, but not in the Shibata et al. results. These differences may be explained in part by the different treatment of the tidal lag angle in the two studies. Whereas Shibata et al. allow a very large lag angle to develop (almost 45°; see Fig. 3e in Shibata et al. 1992a and Fig. 1c in Shibata et al. 1992b), Davies et al. consider this an entirely spurious effect and realign the axes of the two stars by hand just before contact. Clearly, better initial conditions are needed that incorporate self-consistently the effects of the gravitational radiation reaction, such as the tidal lag angle. Extensive tests of the numerical methods are also needed, in particular, to evaluate the magnitude of artificial vorticity generation in extreme shear flows.

This work has been supported by a Hubble Fellowship to F. A. R. funded by NASA through Grant HF-1037.01-92A from the Space Telescope Science Institute, which is operated by the Association of Universities for Research in Astronomy, Inc., under contract NAS5-26555. Partial support was also provided by NSF Grant AST 91–19475 and NASA Grant NAGW–2364 to Cornell University. Computations were performed on the Cornell National Supercomputer Facility, a resource of the Center for Theory and Simulation in Science and Engineering at Cornell University, which receives major funding from the NSF and IBM Corporation, with additional support from New York State and members of its Corporate Research Institute.

Table 1: SUMMARY OF PRINCIPAL RESULTS

| Property | $\Gamma = 2$, $q = 1$ | $\Gamma = 3$, $q = 1$ | $\Gamma = 3$, $q = 0.85$ | $\Gamma = 3$, $q = 0.5$ |
|---|---|---|---|---|
| $r_{dyn}$ | $\approx 3$ | $2.97 \pm 0.02$ | $2.95 \pm 0.05$ | $2.95 \pm 0.05$ |
| Final config. | axisymmetric | triaxial ellipsoid | triaxial ellipsoid | detached binary |
| $M_{halo}/M_{tot}$ | 0.20 | 0.18 | 0.13 | — |
| $\rho_c$ | 0.8 | 0.4 | 0.4 | — |
| $\Omega_c$ | 0.6 | 0.5 | 0.5 | — |
| $(T/|W|)_c$ | 0.13 | 0.14 | 0.14 | — |
| $(r_O/M)h_{max}$ | $2.4 \times (M/R)$ | $2.2 \times (M/R)$ | $1.6 \times (M/R)$ | $0.8 \times (M/R)$ |
| $(r_O/M)h_{final}$ | 0 | $0.2 \times (M/R)$ | $0.03 \times (M/R)$ | $0.6 \times (M/R)$ |
| $L_{max}/L_o$ | $0.55 \times (M/R)^5$ | $0.37 \times (M/R)^5$ | $0.14 \times (M/R)^5$ | $0.018 \times (M/R)^5$ |

---

Fig. 1.— Time evolution of the virial ratio $(2T + 2U + W)/|W|$ during a constant-$r$ relaxation calculation (a) and during a quasi-static scan of the equilibrium sequence (b). This is shown here for two identical polytropes with $\Gamma = 5/3$ (polytropic index $n = 1.5$). At $t = 0$, two *spherical* polytropes are placed in the system. In (a) we keep the separation constant at $r = 2.5$, whereas in (b) $r$ is forced to decrease slowly in time (See text for details). The final configuration obtained in (a) satisfies the virial theorem for hydrostatic equilibrium to an accuracy $\lesssim 10^{-3}$. In (b), all configurations with $2.5 < r < 3.3$ satisfy the virial theorem to within $< 2 \times 10^{-3}$. For $r > 3.3$ ($t < 10$), the two spherical polytropes are still relaxing to their new equilibrium shapes, while for $r < 2.5$ ($t > 90$), the system is too rapidly approaching its mass shedding limit and large deviations from equilibrium become inevitable.

Fig. 2.— Variation of the total angular momentum $J$ along the equilibrium sequence for two identical $n = 1.5$ polytropes considered in Fig. 1. The dots are from different SPH relaxation calculations at constant $r$ like the one depicted in Fig. 1a for $r = 2.5$. The solid line is from the SPH scan of the equilibrium sequence considered in Fig. 1b. For comparison, the dotted line shows the result for two point masses, the short-dashed line for two *spherical* polytropes, and the long-dashed line shows the result obtained in LRS1 for two stars modelled as compressible Darwin ellipsoids. In (b) we show a blow-up of the region near the minimum of $J$ (secular stability limit). Note the excellent quantitative agreement between the fully numerical SPH results and the quasi-analytic approximate treatment of LRS, especially before the minimum of $J$. In contrast, there are large departures from point-mass or even rigid-sphere behavior near the minimum.

Fig. 3.— Dynamical evolution of the binary separation $r$ in three different SPH calculations. The initial conditions are equilibrium configurations for two identical polytropes with $\Gamma = 3$ and initial separations $r = 2.95$, $3.0$, and $3.1$. The configuration with $r = 2.95$ is just beyond the dynamical stability limit.

Fig. 4.— Evolution of the separation $r$ and radial infall velocity $v_r = \dot{r}$ for the unstable binary with initial $r = 2.95$ already considered in Fig. 3. For comparison, the dashed line shows the radial infall velocity corresponding to the gravitational radiation reaction at the same $r$ (calculated in the quadrupole approximation for two point masses).

Fig. 5.— Evolution of the unstable binary with $\Gamma = 3$, mass ratio $q = 1$, and initial separation $r = 2.95$. Projections of all SPH particles onto the orbital $(x, y)$ plane are shown at various times. The orbital rotation is counterclockwise. The initial orbital period $P_{orb} \approx 24$ in our units (defined in §2.2).

Fig. 6.— Time evolution of the various energies during the dynamical coalescence shown in Fig. 5. The system remains close to virial equilibrium throughout its evolution, in spite of the complete reorganization of its internal structure through large amplitude mass motions. The total energy $E_{tot} = T + W + U$ is conserved to a numerical accuracy of about $10^{-2}$.

Fig. 7.— Large-scale view of the spiral arm development during the dynamical coalescence shown in Fig. 5. The outer spiral arms remain narrow and well-defined for a very large number of central dynamical times.

Fig. 8.— Last stable binary configuration (top), and final merged configurations (bottom) for systems with $q = 1$ and $\Gamma = 5/3$ (a), 3 (b), and 10 (c). Contours of constant density are shown in the equatorial plane. The two outermost contours (dashed lines) correspond to $\rho = 0.001$ and $\rho = 0.01$. The inner contours (solid lines) correspond to $\rho = 0.1, 0.2, 0.3, \cdots$, up to the central density $\rho_c$. In (a) $\rho_c \approx 1.2$, in (b) $\rho_c \approx 0.4$, and in (c) $\rho_c \approx 0.3$. These values are all in units of $M/R^3$ (cf. §2.2). Final merged configurations in (b) and (c) are clearly triaxial.

Fig. 9.— Density and velocity profiles inside the merged configuration with $\Gamma = 3$ and $q = 1$ shown in Fig. 8b (solid lines). Here $m$ is the mass contained within a surface of constant density $\rho(m)$. The quantity $\Omega(m)$ shown is the mass-averaged angular velocity inside that surface. The vertical dashed line corresponds to the constant-density surface with maximum equatorial radius $r = 1.5$. It shows the approximate location of the core boundary, where the density drops abruptly by more than one order of magnitude. The dotted lines show the profiles for the merged configuration resulting from the coalescence of a binary with $\Gamma = 3$ and $q = 0.85$ (see §3.2.2).

Fig. 10.— Gravitational radiation waveforms vs retarded time for the dynamical coalescence of the systems with $q = 1$ and $\Gamma = 5/3$, 3, and 10. Here quantities are labeled in geometrized units ($G = c = 1$). Amplitudes of the two polarization states of the radiation are shown for an observer situated at a distance $r_O$ along the rotation axis ($\theta = 0$). After the onset of mass shedding ($t \approx 40$), the amplitude drops abruptly to zero for $\Gamma = 5/3$, whereas it drops to a smaller but finite value for $\Gamma = 3$ and $\Gamma = 10$.

Fig. 11.— Gravitational radiation luminosity corresponding to the waveforms shown in Fig. 10. The solid line is for $\Gamma = 3$, the dashed line for $\Gamma = 5/3$, and the dotted line for $\Gamma = 10$. Geometrized units are used as in Fig. 10, and $L_o \equiv G/c^5$.

Fig. 12.— Evolution of the unstable binary with $\Gamma = 3$, mass ratio $q = 0.85$, and initial separation $r = 2.9$. Conventions are as in Fig. 5.

Fig. 13.— Decomposition of the SPH particles in the final merged configurations with $\Gamma = 3$ according to their binary component of origin. Projections onto the orbital plane are shown. Top frames show all SPH particles coming originally from the more massive component. Middle frames show the particles coming from the less massive component. All particles are combined to produce the plots at the bottom. The left plots are for $q = 1$ (two identical stars, cf. Fig. 5), the right plots for q=0.85 (cf. Fig. 12).

Fig. 14.— Evolution of the unstable binary with $\Gamma = 3$, mass ratio $q = 0.5$, and initial separation $r = 2.9$. Conventions are as in Fig. 5. The plots show the system at successive pericenter and apocenter passages. The final state is a stable, detached binary configuration.

Fig. 15.— Time evolution of the binary separation, total entropy, and component masses during the dynamical evolution shown in Fig. 14. The minimum separation during the last pericenter passage at $t \approx 64$ is outside both the Roche limit and dynamical stability limit.

Fig. 16.— Gravitational radiation waveforms for the dynamical evolution of the systems with $\Gamma = 3$ and $q = 0.85$ and 0.5. Conventiona are as in Fig. 10.